\documentstyle[aps,preprint,epsf,prb]{revtex}
\tightenlines
\begin{document}

\title{Time dependent correlations in marine stratocumulus cloud base height
records}

\author{N. Kitova,$^{1}$   K. Ivanova,$^{2}$   M. Ausloos,$^3$   T.P.
Ackerman,$^4$ and M.A. Mikhalev$^1$ }

\address{ $^{1}$Institute of Electronics, Bulgarian Academy of Sciences, 72
Tzarigradsko chaussee, Sofia 1784, Bulgaria \\ $^{2}$Department of Meteorology,
Pennsylvania State University, University Park, PA 16802, USA \\ $^3$SUPRAS, B5,
University of Li$\grave e$ge, B-4000 Li$\grave e$ge, Belgium\\ $^4$Pacific
Northwest National Laboratory, Richland, WA 99352, USA }



\maketitle

\begin{abstract} The scaling ranges of time correlations in the cloud base height
records of marine boundary layer stratocumulus are studied applying the Detrended
Fluctuation Analysis statistical method. We have found that time dependent
variations in the evolution of the $\alpha$ exponent reflect the diurnal dynamics
of cloud base height fluctuations in the marine boundary layer. In general, a
more stable structure of the boundary layer corresponds to a lower value of the
$\alpha$ - indicator, i.e. larger anti-persistence, thus a set of fluctuations
tending to induce a greater stability of the stratocumulus. In contrast, during
periods of higher instability in the marine boundary, less anti-persistent (more
persistent like) behavior of the system drags it out of equilibrium,
corresponding to larger $\alpha$ values. From an analysis of the frequency spectrum, the
stratocumulus base height evolution is found to be a non-stationary process with
stationary increments. The occurrence of these statistics in cloud base height
fluctuations suggests the usefulness of similar studies for the radiation
transfer dynamics modeling. \end{abstract}

 \newpage

\section{Introduction}

Cloud base height (CBH) profiles are difficult object to describe and analyze.
The base of a cloud is determined by the local condensation level up to which
unsaturated air parcels rise. CBH has a complex structure, that varies with the
coordinates of the observation point and time. Such a complexity is due to a
variety of processes that take place in the marine boundary layer (MBL) formed
over the sea surface. Some of these processes are, e.g., turbulent motions,
entrainment,\cite{turner,lilly} radiative transfer and changes in the cloud microphysical
structure. Thus the CBH looks like an erratic quantity in time and space.
Therefore, advanced methods of computational statistical analysis are required in
order to extract meaningful physical information from data series with such a
high level of  complexity that traditional analysis techniques fail to provide.
The detrended fluctuation analysis (DFA) method is designed to search for
long-range correlations\cite{masta} in non-stationary, highly fluctuating signals. The
method has been successfully applied in the investigation of turbulence,\cite{fxnature} DNA
sequences,\cite{DNADFA} ionic transport trough cell membrane,\cite{cell,cellma} foreign currency
exchange rates,\cite{fx1,fx2,fx3} meteorological phenomena,\cite{soimaki,bunde,talkner} and stratus clouds
dynamics.\cite{kimaeeta,plov} Using this technique\cite{DFA} we hereby sort out correlations and
anti-correlations in stratocumulus cloud base height fluctuation data records.
Our findings of the time evolution of the local $\alpha$-exponent (which is the Hurst
exponent\cite{Hu4}),
i.e. describing a scaling law over a certain time range, are related to the
physical processes in the cloud. These results reflect the ability of the method
to indicate the type of physics that underlies the CBH profile phenomenon.

\section{Marine Boundary Layer - structure and processes}

A variety of physical processes takes place in the atmospheric boundary layer. At
time scales of less than one day, significant fluxes of  heat, water vapor and
momentum are exchanged due to entrainment, radiative transfer, and/or turbulence.\cite{garratt,andrews}
The turbulent character of the motion in the atmospheric boundary layer
(ABL) is one of its most important features. The turbulence\cite{frisch} can be caused by
a variety of processes, like thermal convections, or mechanically generated by
wind shear, or following interactions influenced by the rotation of the Earth.\cite{atmoturbulence,turbu}
This complexity of physical processes and interactions between them
create a variety of atmospheric formations. In particular, in a cloudy ABL the
radiative fluxes produce local sources of heating or cooling within the
mixed-layer and therefore can greatly influence its turbulent structure and
dynamics. Moreover the variations in the turbulent structure and dynamics of the
clouds cause subsequent changes in the cloud boundaries, especially in the cloud
base.

The atmospheric boundary layer is characterized by an inner (surface) layer\cite{garratt,abl}
at heights above the aerodynamic roughness length\cite{brud} and below one
tenth of the depth of the ABL, while the outer (Eckman) layer is at higher
distances within the ABL. The outer region of an unstably stratified ABL is often
referred to as the mixed layer\cite{garratt,abl} because of the dominating convective
motions that take place there, generated by strong surface heating from the Sun
or by cloud-topped radiative cooling processes.\cite{andrews} In contrast, a stably
stratified ABL occurs mostly at night in response to the surface cooling due to
long-wave length radiation emitted into the space.

The top of the boundary layer in convective conditions is often well defined by
the existence of a stable layer (capping or subsidence inversion), beneath which
clouds form, the so-called cloud-topped boundary layer (CTBL).\cite{world} In presence
of clouds (shallow cumulus, stratocumulus (Sc) or stratus (St)) the structure of
the ABL is modified because of the radiative fluxes. Phase changes become more
important. During cloudy conditions one can distinguish mainly : (i) the case in
which the cloud and the subcloud layers are fully coupled; (ii) two or more cloud
layers beneath the inversion, with the lower layer well-mixed with an upper
elevated layer, decoupled from the surface mixed layer or (iii) a radiatively
driven elevated mixed cloud layer, decoupled from the surface.

An  interesting case is that of the marine boundary layer characterized by high
concentration of moisture. It is wet, mobile and has a well expressed lower
boundary. The competition between the processes of radiative cooling, entrainment
of warm and dry air from above the cloud and turbulent buoyancy fluxes determine
the state of equilibrium of the cloud-topped marine boundary layer.\cite{garratt} The
Atmospheric Stratocumulus Transition Experiment (ASTEX) was designed to clarify
the transition from stratocumulus to trade cumulus clouds\cite{garratt} in the MBL in the
region of the Azores Islands.\cite{mbl} Several papers have been dedicated to the
relevance of the main processes causing this transition.\cite{astex1,astex2} In Ref. [27] 
Betts et al. found a relationship between the thermodynamic structure of the marine
boundary layer and the diurnal variation of the cloudiness.

\section{DFA method and spectral analysis}

A method that relaxes the requirement of stationarity of the investigated signal
is the detrended fluctuation analysis (DFA) method. The DFA method is a tool used
for sorting out {\it long range correlations} in a non-stationary self-affine
time series with stationary increments. The method has been used previously in
the meteorological field.\cite{fx3,bunde,kimaeeta} It provides a simple quantitative parameter
- the scaling exponent $\alpha$, which is a signature of the correlation
properties of the signal. The Detrended Fluctuation Analysis technique consists\cite{DFA}
in dividing a random variable sequence $y(n)$ of length $N$ into $N/t$
(non-overlapping) boxes, each containing $t$ points ($N/t$ = 4,5, ...). Then, 
the trend (assumed to be linear in this investigation, but it can
be generalized without any difficulty\cite{fx2}) $z(n)=an+b$ in each box is computed
using a linear least-square fit to the data points in that box. The detrended
fluctuation function $F^2(t)$ is then calculated following :

\begin{equation} F^2(t) = {1 \over t } {\sum_{n=kt+1}^{(k+1)t} {\left[y(n)-
z(n)\right]}^2} \qquad \mbox{for} \qquad
k=0,1,2,\dots,\left(\frac{N}{t}-1\right). \end{equation}

Averaging $F^2(t)$ over the $N/t$ intervals gives the mean-square fluctuations

\begin{equation} <F^2(t)>^{1/2} \sim t^{\alpha}. \label{dfa} \end{equation}

If the $y(n)$ data are random uncorrelated variables or short range correlated
variables, the behavior is expected to be a power law\cite{addison} with an exponent\cite{DFA}
$\alpha$=1/2  if the fluctuations are not correlated. An exponent $\alpha$
$\neq$ 1/2 in a certain range of $t$ values implies the existence of long-range
correlations in that time interval as, for example, in fractional Brownian motion.\cite{addison}
A small value of $\alpha$ indicates antipersistence\cite{addison} of correlations,
i.e. a positive fluctuation is more likely to be followed by a negative one than
a positive one. An $\alpha$ value between 0.5 and 1 indicates persistence of
correlations, i.e. a positive fluctuation is more likely to be followed by a
positive one than a negative  one. The $\alpha$-exponent value that holds true 
for a certain time interval called
{\it scaling range}, is a characteristic of the correlations in the
fluctuations of a signal $y(t)$ defined between the beginning $t_0$ and end of
the observations $t_M$, i.e. $[t_0,t_M]$.
The value can vary in time if the process is not stationary and/or the scaling
range is finite (see Sect.6).  The $\alpha$ exponent is related to the usual
fractal dimension\cite{addison,malamud} $D$.

The concept of the invariance of the fractal shapes\cite{addison} at different scales
(self-similarity) can be transposed in the statistical analysis of a time series.
One searches for a given scale range where the statistical properties of the
signal are invariant. By definition\cite{addison} the time series is self-affine if its
spectrum has a power law dependence.

\begin{equation} S(f) \sim (1/f)^{\beta}. \label{Sf} \end{equation}

It has been shown by Heneghan and McDarby\cite{heneghan} that the relationship
$\beta=1+2\alpha$ holds true for stochastic processes, i.e. for fractional
Brownian walks. Depending on the value of the spectral exponent $\beta$, one has
the following cases\cite{heneghan,davis} : if $\beta<1$, the process is stationary; if
$\beta>1$, the process is non-stationary; if $1<\beta<3$ the process is
non-stationary with stationary increments. In the case of white noise $\beta=0$
while $\beta=2$ for a Brownian walk signal.

\section{Data}

Cloud base height data has been obtained during the Atlantic Stratocumulus
Transition Experiment (ASTEX) in June 1992, at the Azores Islands using a laser
ceilometer. The wavelength of the vertically transmitted into the atmosphere
laser beam is 0.904 mm. The ceilometer works with a 15 m spatial resolution and
reaches the maximum measurable height of 4 km. The time resolution of  CBH
records is 30 seconds. The cloud base height values were selected from the
profiles of the backscattered signal intensity, using manufacturer's algorithm
based on a voltage threshold for the maximum value of the backscattered radiation
at a height\cite{thesis} giving the local condensation level up to which unsaturated air
parcels rise. We use data records measured on June 15, 1992 (Fig. 1(a)) and June
18, 1992 (Fig. 2(a)) to determine (i) the scaling properties of CBH fluctuations;
(ii) to classify the type of time correlations, (iii) to investigate the diurnal
changes in the CBH evolution, and (iv) to identify the type of the time dependent
long-range correlation evolution.

According to,\cite{betts} on June 15, 1992 an anticyclone was centered north of Santa
Maria - one of the Azores Islands, where the measurements were taking place,
producing a strong subsidence inversion beneath which were observed marine
boundary layer clouds. The nocturnal MBL was found to be well-mixed, generally
coupled to the surface, with a solid layer of stratocumulus, while the daytime
MBL was observed to be decoupled from the surface, because of the coexistence of
a stratocumulus cloud layer and a cumulus marine subcloud layer. After the
sunrise the MBL became intermittently decoupled and CBH varied dramatically as a
cumulus formed at the top of the marine subcloud layer. In this decoupled MBL,
the cumulus convection, that arises from the latent instability generated in the
marine subcloud layer, acts toward a recoupling of the formally decoupled cloud
layers that persist beneath the trade inversion. Five sequences of strong
decoupling were observed that day (Fig. 1(a)).

For June 18, 1992 we do not have any information about the meteorological
situation, but in analogy with data in Fig. 1(a), the CBH profile on Fig. 2(a)
leads one to distinguish periods of increased variability of CBH, followed by
short CBH fluctuations.

\section{Scaling properties of the CBH data}

The CBH evolution, reported in Figs. 1(a) and 2(a) through $S(f)$, can be
regarded as a non-stationary process with stationary increments, since their
spectral exponents are $\beta=1.28\pm0.1$, for June 15, 1992 (Fig. 3) and
$\beta=1.49\pm0.08$, for June 18, 1992  (Fig. 4). The data error bars are
calculated from the r.m.s deviations in the fit according to standard techniques
and have their $95\%$ confidence interval conventional meaning.\cite{statbook}

The type of correlations of the CBH fluctuations can be probed using the
Detrended Fluctuation Analysis (DFA) method. Applying the DFA method, we find
that long-range time correlations in CBH fluctuations exist and are of an
anti-persistent type with  $\alpha=0.24\pm0.002$ for
June 18, 1992 and $\alpha=0.21\pm0.005$ for June 15, 1992 (Fig. 5). They hold
for a time interval approximately equal to 140 min in the first case and 40 min
for the second. In this scaling range, the relationship $\beta=1+2\alpha$ is
reasonably satisfied. The flat like regions of both curves (Fig. 5) correspond to
low values of the Hurst exponent  $\alpha$ ($\alpha=0.06\pm0.003$ for June
15, and $\alpha=0.10\pm0.003$ for June 18) and indicate the presence of
$1/f$-like noise.

For completeness, we show in Fig. 6, the distribution function of the CBH
fluctuations for the 30 sec time lag interval. Notice the similarity between both
cloud cases and the marked departure from the Gaussian behavior, as numerically
exemplified in the $S(f)$ and DFA correlation methods in fact. A similar double
triangular pyramid shape has been observed in other meteorological cases.\cite{genaachen}

\section{Time correlation analysis }

It is of interest to test if the correlations remain the same for shorter
intervals accommodated into $[t_0,t_M]$ or if they change with time, as should be
anticipated for nonstationary time series data. In order to probe the existence
of so called {\it locally correlated} and {\it decorrelated} sequences, one can
construct an ''observation box'' with a certain width, $\omega$, of size 7h,
place the box at the beginning of the data, calculate $\alpha$ for the data in
that box, move the box by $\Delta\omega$= 30 min (or 60 points)  toward the right
along the signal sequence, calculate $\alpha$ in that box, a.s.o. up to the
$N$-th point of the available data. A time dependent $\alpha$ exponent may be
expected, and is given at the end of each box, thus for $t$ ranging from $\Delta
\omega$ to $N$.

Thus we obtain an "instantaneous measure" for the degree of long-range
correlations. Results for this instantaneous $\alpha$-exponents are shown in
Figs. 1(b) and 2(b) for the cloud base height data on June 15, 1992 and June 18,
1992. Note that each $\alpha$ value represents the behavior resulting from all
points in the box, which continuously overlap. This approach seems suitable to
cloud base height data because it can be expected to reveal changes in the
correlations dynamics  of the clouds at various times for a given time lag
$\Delta\omega$.

The $\alpha$ evolution is within the limits of a quite anti-persistent behavior
with a transition from weak to strong, at various instant, in other words,
anti-persistence is found in the underlying dynamics. Five regions in the CBH
evolution, corresponding to the description given in section 2, can be
distinguished\cite{DFA} for the CBH data measured on June 15. The first region lasts
from 0 to 6 h, the second from 6 to 13 h, the third holds from 13:00 to 16:00 h,
the fourth - from 16:00 to 20:00 h and the fifth one from 20:00 to 24:00 h. The
transition occurring from 08:00 to 10:30 in the evolution of the $\alpha$
exponent indicates the increasing role of a process that generates the second
structure in the CBH that appears between 10-13 h in Fig. 1(b). This process is
characterized by a mean value of the local $\alpha=0.24\pm0.01$ and is
understood as a process of $destabilization$, in the {\it decoupled MBL}. Between
14-18 h the downward trend of the $\alpha$  exponent indicates the increasing
dominance of a process that generates the third part of CBH evolution that exists
between 17-19 h in Fig. 1(b). The mean value of the local a exponent is $\alpha$
=0.20$\pm$0.02. From 19 h to 20:30 h a very slow upward trend leads to the
process of the fourth region between 21:30 h - 24 h in Fig. 1(b), with a mean
local $\alpha$ =0.26$\pm$0.02.

For June 18 we distinguish three main regions in the CBH evolution: 0-10 h, 10-19
h and 19-24 h. The trends in the $\alpha$  exponent evolution are as follows:
0-10 h - process, generating the first CBH structure, with mean local $\alpha$
=0.17$\pm$0.02. This corresponds to the dynamics of the MBL before its
decoupling. For the period 10-16 h a large upward transition leads to a weak
anti-persistent correlation in CBH fluctuations with a mean value of the local
$\alpha$ =0.35$\pm$0.02. Hence the dominant process is destabilizing the MBL
structure. For the time 16-20 h the dynamics of the process changes again,
characterized by a mean local $\alpha$ =0.26$\pm$0.01 and a return to more stable
MBL structure.

Thus the evolution of the Hurst exponent $\alpha$ in the cases that we
have investigated, consists in the successive transitions from lower to higher
values and vice versa. This corresponds to transitions from strong to weak
anti-persistent correlations and reflects the changes in the underlying dynamics
of the MBL.

Notice that the DFA function being a measure of the root mean square deviation
from the linear best fit (a ''local trend'') is therefore a measure of a
height-height correlation function. The so-called height is nothing else than a
mass threshold. Therefore this DFA function informs on a corresponding physical
property, i.e. the local compressibility (or bulk modulus) of the cloud fluid.\cite{phys,statphys}

\section{Conclusions}

We have presented an analysis of a marine stratocumulus cloud base height data
series using the DFA method and the spectral density method in order to
investigate the scaling range of time correlations in the CBH fluctuations
searching for scaling laws and persistence effects. In all cases anti-persistent
correlations hold with a well established day-night evolution measured by the
instantaneous $\alpha$ index , - itself characterizing the dominant physical
process. The transition, revealed by the trend in the $\alpha$ evolution, between
processes inducing different levels of anti-persistent correlations of the
fluctuations could be used as a trace parameter of the CBH dynamics. The accuracy
of these statistics in CBH fluctuations suggests the usefulness of similar
studies for the radiation transfer dynamics.

The DFA method allows one to obtain the scaling ranges and crossovers of the
correlations between the fluctuations of a signal.  The $\alpha$ exponent is in fact
directly related to that for the auto correlation function.\cite{cell} Incidently, if
the relationship $\gamma = 2(1-\alpha)$ is supposed to hold for describing the
power law dependence of the autocorrelation function, a value $\gamma$ = 1.56 is
expected.  This means that the CBH fluctuations can be empirically represented by
a fractional Brownian motion process.\cite{addison} The oscillations between different
stability regimes could thus be mapped onto a Kramer problem,\cite{statphys} itself
mimicking the basic Langevin equation for the signal increments.

Nevertheless this knowledge of the second moment of a signal is not usually
enough in order to have a full understanding of the physical process nor of its
underlying physical mechanisms governing the CBH. It would be of interest to
couple these results to some analysis of the local (vertical and horizontal)
velocity fluctuations, and of the temperature fluctuations in order to obtain
additional information, whence the viscosity and thermal conductivity, for a
better understanding of cloud physics and  weather predictability.

\section{Acknowledgments}

This research was partially supported by Battelle grant number 327421-A-N4.

\newpage

\begin{figure}[ht] \begin{center} \leavevmode \epsfysize=8cm
\epsffile{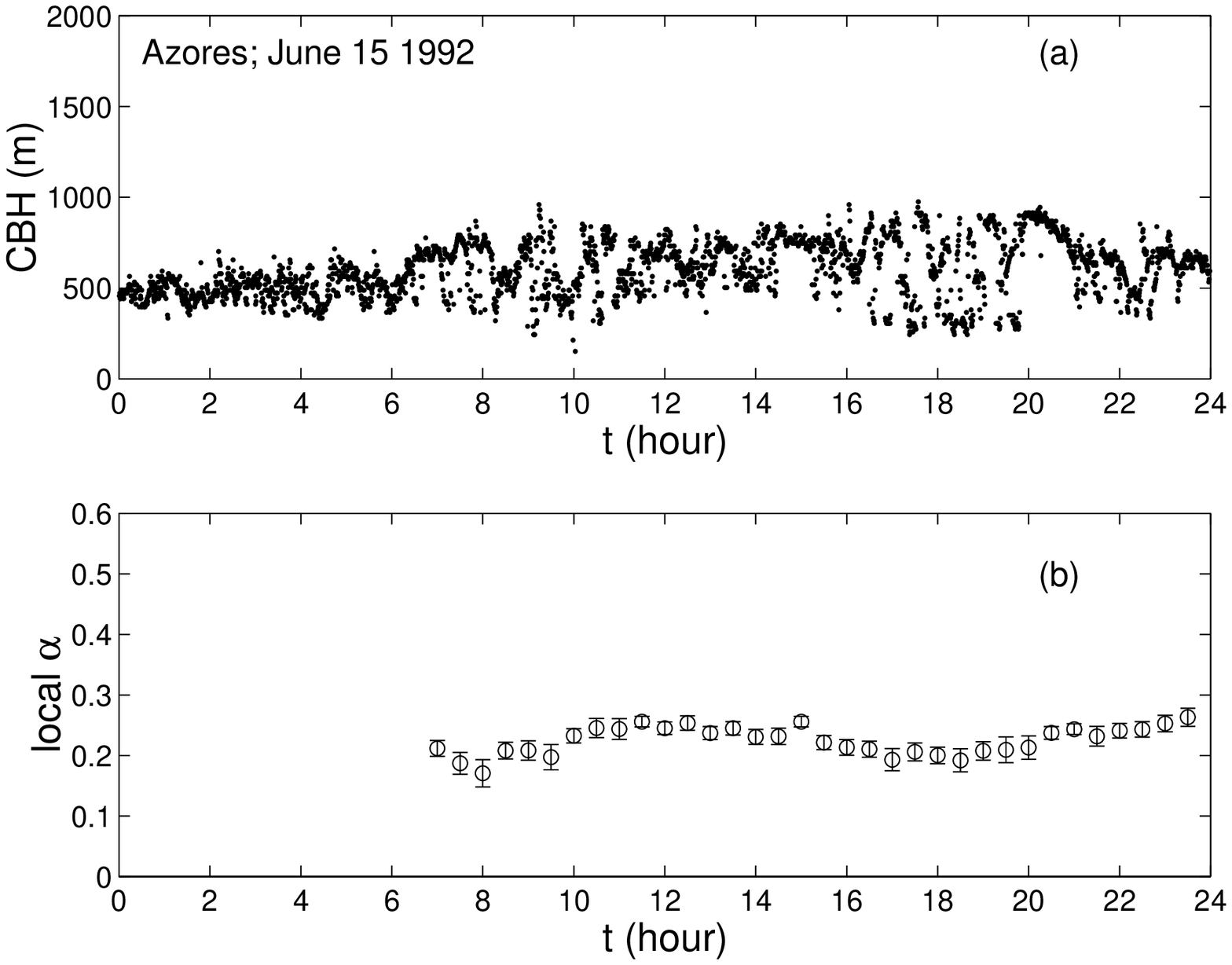} \caption{(a) CBH fluctuations for June 15, 1992; (b) The $\alpha$ evolution,
clarifying the transitions between strong and weak anti-persistent behavior of
CBH fluctuations.}
\end{center}
\end{figure}

\begin{figure}[ht] \begin{center} \leavevmode \epsfysize=8cm
\epsffile{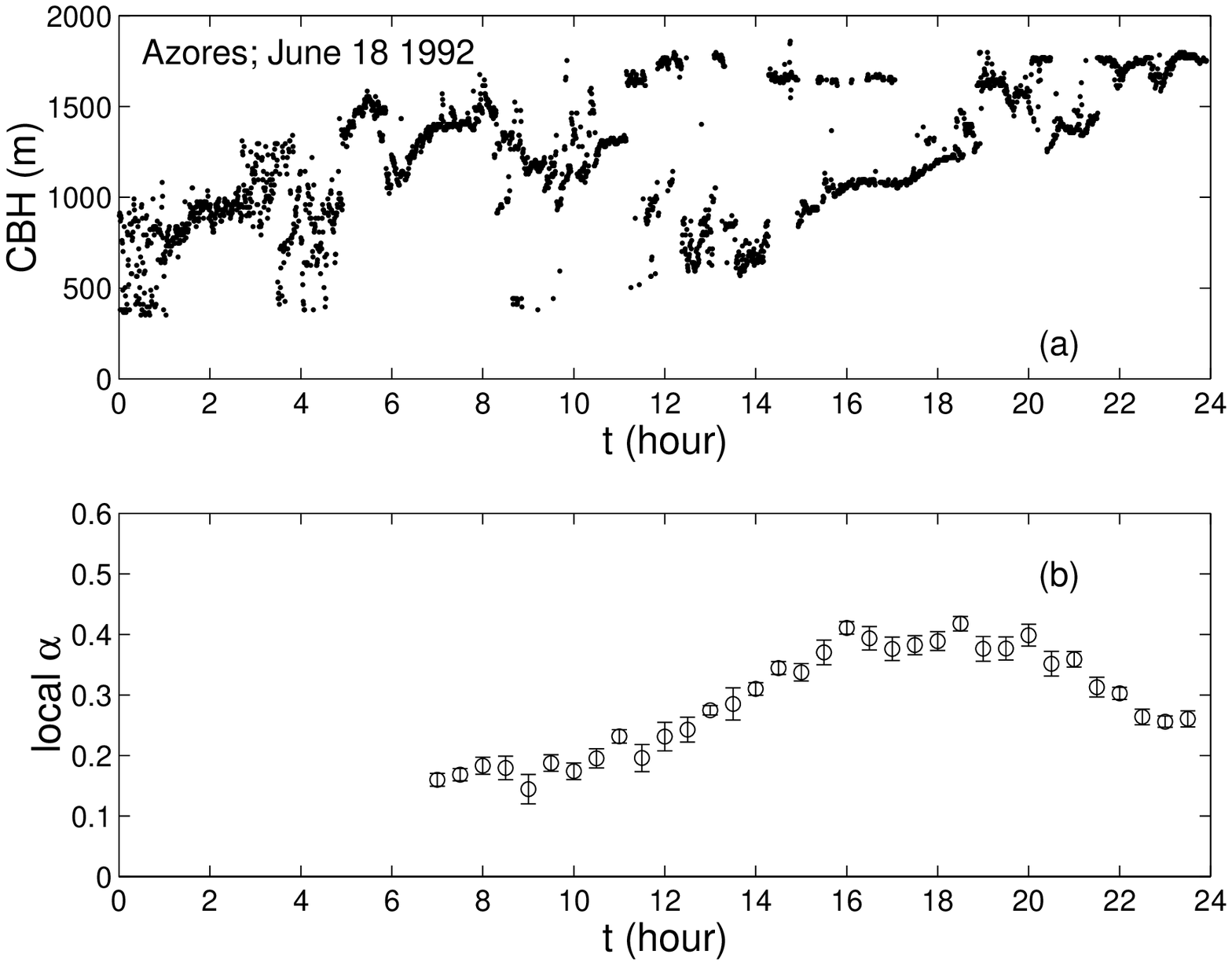} \caption{(a) CBH fluctuations for June 18, 1992; 
(b) The $\alpha$  evolution,
clarifying the transitions between strong and weak anti-persistent behavior of
CBH fluctuations.}
\end{center}
\end{figure}

\begin{figure}[ht] \begin{center} \leavevmode \epsfysize=8cm
\epsffile{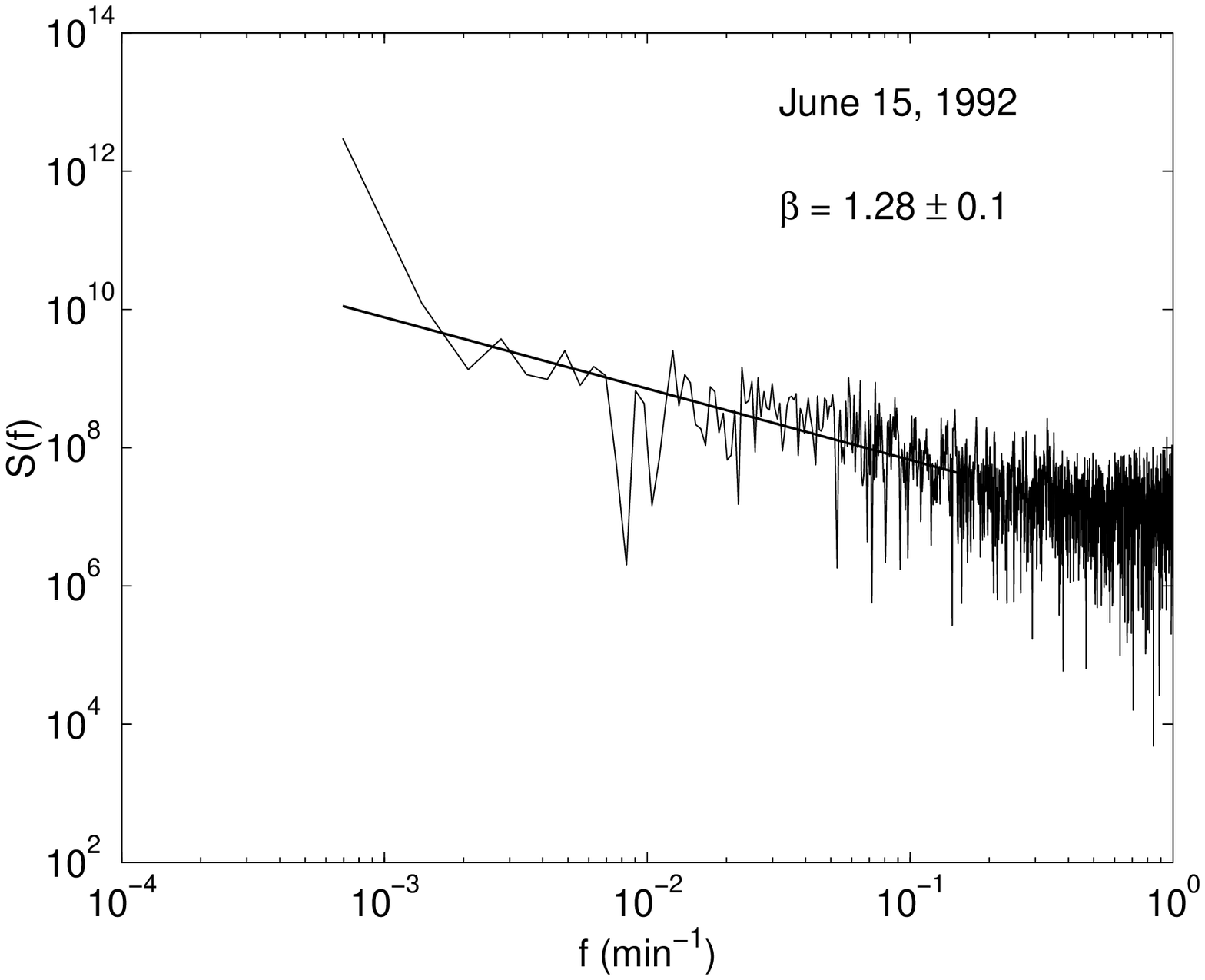} \caption{
The power spectrum of the cloud base height data (data in Fig. 1(a)). A
spectral exponent $\beta$ =1.28$\pm$0.1 characterizes the correlations of
fluctuations.}
\end{center}
\end{figure}

\begin{figure}[ht] \begin{center} \leavevmode \epsfysize=8cm
\epsffile{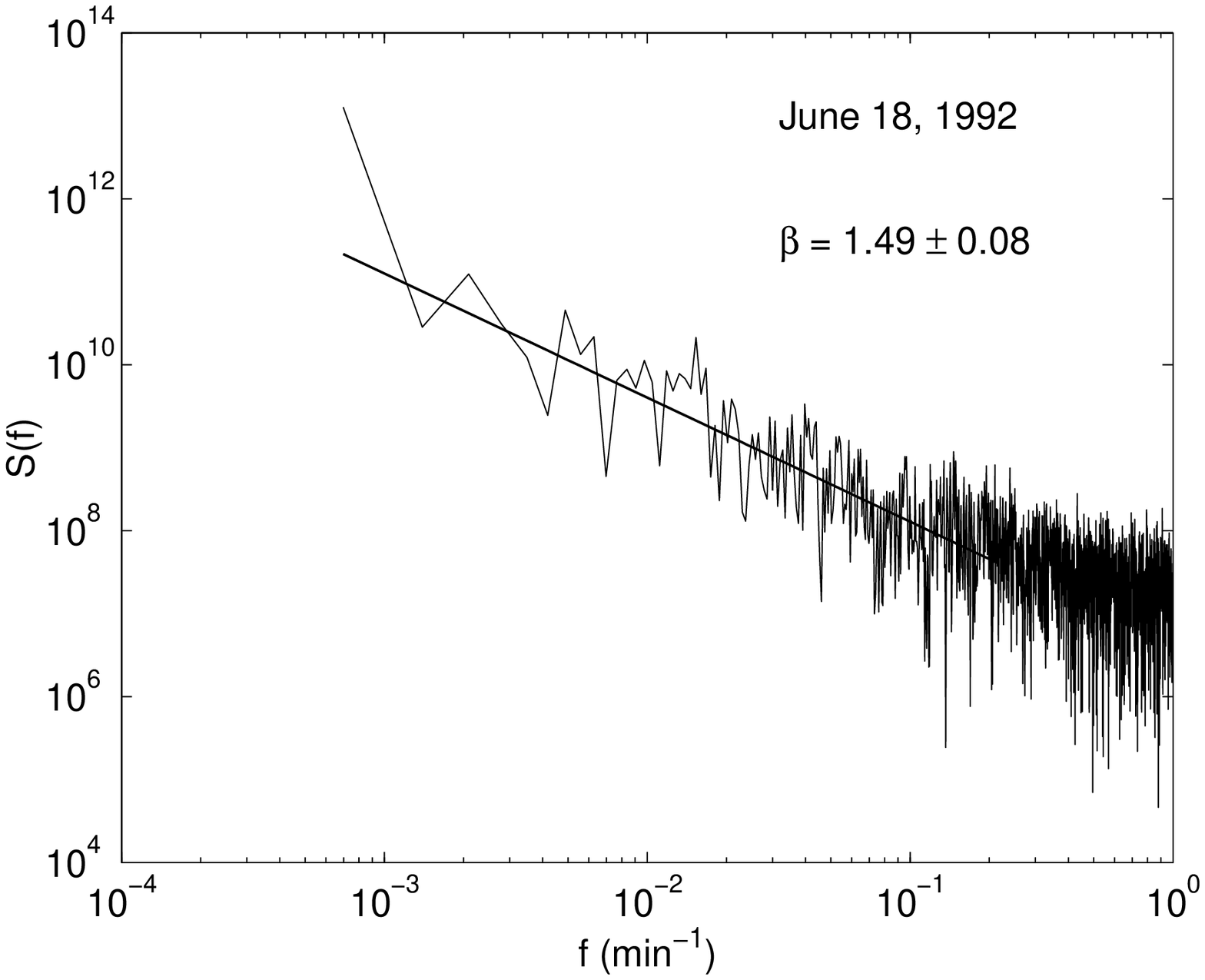} \caption{
The power spectrum of the cloud base height data (data in Fig. 2(a)). A
spectral exponent  $\beta$=1.49$\pm$0.08 characterizes the correlations of fluctuations.}
\end{center}
\end{figure}

\begin{figure}[ht] \begin{center} \leavevmode \epsfysize=8cm
\epsffile{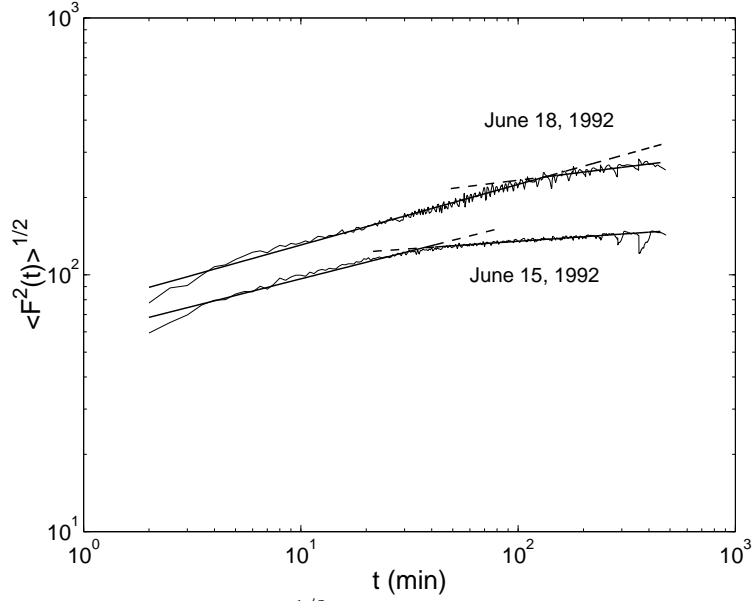} \caption{ The DFA function $<F(t)>^{1/2}$ for the CBH data
measured on June 15, 1992 (data in Fig. 1(a)). Scaling properties change from $\alpha_1=0.21\pm0.005$
to $\alpha_2=0.06\pm0.003$ for time lags longer than 40~min. CBH data measured on June 18, 1992
(data in Fig. 2(a)) scale with $\alpha_1=0.24\pm0.002$ for time lags smaller than 140~min
and $\alpha_2=0.10\pm0.003$ after that.}
\end{center}
\end{figure}

\begin{figure}[ht] \begin{center} \leavevmode \epsfysize=8cm
\epsffile{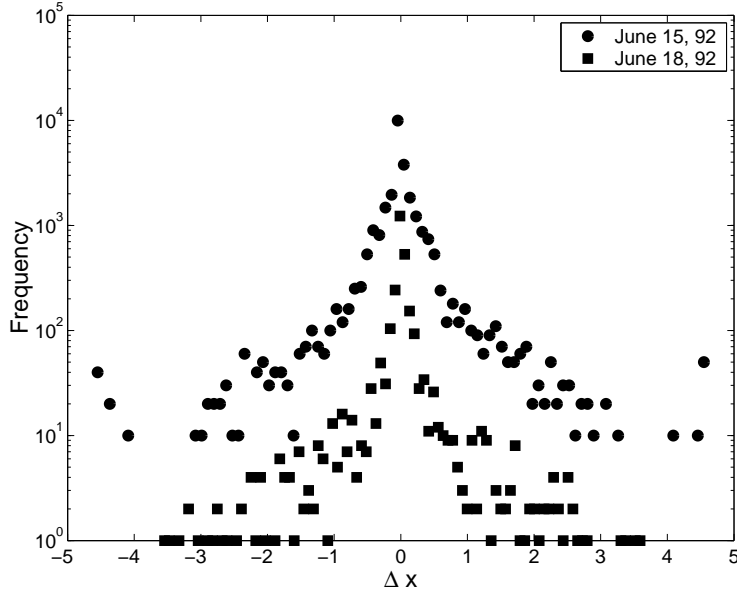} \caption{ Partial distribution function (PDF) of the
CBH signal increments (discretization step = 30~s) 
for both June 15 (black dots) and June 18, 1992
(black squares)  stratocumulus clouds observed on Azores Islands. The x-axis is
normalized in units of the standard deviation of the corresponding PDF for a time
lag 960 s,  - supposedly long enough such that  each PDF then corresponds to a
Gaussian distribution.  The standard deviation for June 15 is 198.9 m and the one
for June 18 is 323.4 m. The June 15 data has been displaced vertically by one
decade.}
\end{center}
\end{figure}

\end{document}